\documentclass[fleqn,twoside]{article}
\usepackage{espcrc2}

\usepackage{graphicx}
\usepackage[figuresright]{rotating}
\usepackage{amsmath}

\def\gsim{\raise0.3ex\hbox{$>$\kern-0.75em\raise-1.1ex\hbox{$\sim$}}}
\def\lsim{\raise0.3ex\hbox{$<$\kern-0.75em\raise-1.1ex\hbox{$\sim$}}}

\title{AUGER-HiRes results and models of Lorentz symmetry violation}

\author{Luis Gonzalez-Mestres\address{LAPP, Universit\'e de Savoie, CNRS/IN2P3, B.P. 110, 74941 Annecy-le-Vieux Cedex, France}}

\begin{document}

\begin{abstract}
The implications of AUGER and HiRes results for patterns of Lorentz symmetry violation (LSV) are examined, focusing on weak doubly special relativity (WDSR). If the Greisen-Zatsepin-Kuzmin (GZK) cutoff is definitely confirmed, the mass composition of the highest-energy cosmic-ray spectrum will be a crucial issue to draw precise theoretical consequences from the experimental results. Assuming that the observed flux suppression is due to the GZK mechanism, data will allow in principle to exclude a significant range of LSV models and parameters, but other important possibilities are expected to remain open : Lorentz breaking can be weaker or occur at a scale higher than the Planck scale, unconventional LSV effects can fake the GZK cutoff, threshold phenomena can delay its appearance…  Space experiments appear to be needed to further test special relativity. We also examine the consequences of AUGER and HiRes data for superbradyons. If such superluminal ultimate constituents of matter exist in our Universe, they may provide new forms of dark matter and dark energy. 
\vspace{1pc}
\end{abstract}

\maketitle

\section{Introduction}

Data \cite{AugerData1,HiresData1} and recent analysis \cite{AugerData2,HiresData2} from the AUGER \cite{AUG} and HiRes \cite{Hires} collaborations possibly confirm the existence of the Greisen-Zatsepin-Kuzmin (GZK) cutoff \cite{GZK1,GZK2} for ultra-high energy cosmic rays (UHECR). But alternative scenarios are still possible and, even assuming that the observed effect is due to the GZK cutoff, the theoretical interpretation and consequences will also depend on other features of the experimental results. A crucial information is the composition of the highest-energy cosmic-ray spectrum \cite{Gonzalez-Mestres2008}. 

It had been stated \cite{AugerData3} that cosmic rays at energies above 6 x $10^{19} ~ eV$ appeared to be protons, mostly from the supergalactic plane. But this conclusion was challenged by Gorbunov et al. \cite{Gorbunov} and by Fargion \cite{Fargion}, who suggested that the most energetic observed cosmic rays are heavier objects from extragalactic sources like the nearby radio galaxy Centaurus A. Dermer \cite{Dermer} favors nuclei of the CNO group. Stanev \cite{Stanev} finds correlation with the updated supergalactic plane, but tends to confirm the role of Cen A. Gureev and Troitsky \cite{Gureev} get results compatible with intermediate mass nuclei accelerated in Cen A. In recent papers, the AUGER collaboration has concluded \cite{AugerData2} that the highest-energy cosmic rays are intermediate between protons and Fe nuclei. More statistics is needed and, eventually, separate identification of fluxes for nucleons and for different kinds of nuclei. In \cite{Gonzalez-Mestres2008}, we briefly illustrated the relevance of this question for models of Lorentz symmetry violation that can in principle produce observable effects in the GZK energy region. 

Patterns of standard (strong) doubly special relativity \cite{SDSR} (SDSR), where the laws of Physics are exactly the same in all inertial frames, do not lead to any testable new physics in the GZK region. That LSV models without a privileged inertial frame cannot reproduce a possible absence of the GZK cutoff was first noticed in \cite{gon02} looking in detail at the numerical predictions of the Kirzhnits-Chechin model \cite{Kir}. 

Instead, the WDSR pattern considered in our papers since 1995 \cite{gona,gonb} can lead to observable effects for UHECR. Contrary to SDSR, the dynamics incorporates a fundamental length scale, a privileged (vacuum) rest frame and, possibly, a new space-time symmetry for the ultimate constituents of matter (superbradyons \cite{gonSL} ?). Energy and momentum remain additive for free particles and energy-momentum conservation is preserved, but standard particles obey a deformed relativistic kinematics (DRK) that cannot be made equivalent to the conventional Lorentz symmetry. The laws of Physics are not identical in all inertial frames, but relativity remains a low-energy limit for standard matter in the vacuum rest frame (VRF). Special relativity is recovered if the fundamental length is set to zero. Data from AUGER and HiRes, and from other UHECR experiments, can test WDSR models but not SDSR. 

We consider here a simple pattern of quadratically deformed relativistic kinematics (QDRK), proposed in our april 1997 paper \cite{gon97a} and developed in subsequent articles \cite{gona,gonb}. As emphasized in these papers, simple realizations of linearly deformed relativistic kinematics (LDRK, see \cite{gona,gonb}) appear to be clearly excluded by data. LDRK (QDRK) models are those where the effective parameter of Lorentz symmetry violation in the equation relating energy and momentum varies linearly (quadratically) with the particle momentum at high energy. QDRK generated at the Planck scale is not ruled out by the AUGER - HiRes data, and can even produce effects (like spontaneous photon emission at ultra-high energies) able to fake the GZK cutoff.

Possible thresholds changing the energy dependence of LSV parameters \cite{gonb} are not dealt with here. If they exist below the Planck scale, a wider range of models will escape experimental bounds from AUGER and HiRes. Another open question is that of the validity of standard interaction models for UHECR with LSV, as in this case a particle at ultra-high energy (UHE) is not the same physical object as the particle at rest. Furthermore, even assuming special relativity to hold, present extrapolations of hadronic interaction models to the highest observed cosmic-ray energies present uncertainties that can influence the physical interpretation of data.

Potential implications for the superbradyon hypothesis \cite{gonSL} are also considered.
Such superluminal preons may be the ultimate constituents of matter, as well as sources of UHECR and of large-scale phenomena. DRK naturally emerges in scenarios where particles are excitations of more fundamental matter with a new length scale. Superbradyons with speed larger than $c$ (speed of light) can spontaneously emit conventional particles. A cosmological sea of superbradyons with speeds  $\simeq ~ c$ may thus have been generated \cite{Gonzalez-Mestres2008}.

\section{QDRK}

Assuming the existence of an absolute local rest frame (the VRF) for conventional matter, together with a fundamental length scale $a$ where new physics can appear \cite{gon97a}, QDRK is given in the VRF by an equation of the form \cite{gon97a,gon97b}:
\begin{equation}
E~=~~(2\pi )^{-1}~h~c~a^{-1}~e~(k~a)
\end{equation}
$h$ being the Planck constant, $c$ the speed of light, $k$ the wave vector, and $[e~(k~a)]^2$ a convex function of $(k~a)^2$ obtained from vacuum dynamics. Expanding (1) for $k~a~\ll ~1$ , we get \cite{gon97b}:
\begin{equation}
\begin{split}
e~(k~a) ~ \simeq ~ [(k~a)^2~-~\alpha ~(k~a)^4~ \\
+~(2\pi ~a)^2~h^{-2}~m^2~c^2]^{1/2}
\end{split}
\end{equation}
where $p$ is the particle momentum, $\alpha $ a model-dependent constant and {\it m} the mass of the particle. For $p~\gg ~mc$ , one has:
\begin{equation}
E ~ \simeq ~ p~c~+~m^2~c^3~(2~p)^{-1}~ \\
-~p~c~\alpha ~(k~a)^2/2~~~~~
\end{equation}

The Earth is assumed to move slowly with respect to the VRF, so that VRF equations for UHECR apply in the laboratory rest frame to a good approximation. The "quadratic deformation" approximated by $\Delta ~E~=~-~p~c~\alpha ~(k~a)^2/2$ in (3) implies a Lorentz symmetry violation in the ratio $E~p^{-1}$ varying like $\Gamma ~(k)~\simeq ~\Gamma _0~k^2$ where $\Gamma _0~ ~=~-~\alpha ~a^2/2$ . If $c$ is a universal parameter for all particles, the QDRK defined by (1) - (3) preserves Lorentz symmetry in the limit $k~\rightarrow ~0$. In terms of the equivalent fundamental energy scale $E_a ~ =~ h ~ c ~(2 ~\pi ~a)^{-1}$, equation (3) becomes :
\begin{equation}
\begin{split}
E ~ \simeq ~ p~c~+~m^2~c^3~(2~p)^{-1}~ \\
-~p~c~\alpha ~(p~ c ~E_a^{-1})^2/2~~~~~
\end{split}
\end{equation}
and $\Delta ~E~=~-~p~c~\alpha ~(p~ ~c ~E_a^{-1})^2/2$.

\subsection{Critical energy scales}

Above the transition energy scale $E_{trans}~$ $\approx ~\alpha ^{-1/4} ~ (h ~ m)^{1/2} ~ c^{3/2} ~ (2 ~ \pi ~ a)^{-1/2}$ = $(m ~ E_a)^{1/2}~ c~\alpha ^{-1/4}$, the deformation $\Delta ~E$ dominates over the mass term $m^2~c^3~(2~p)^{-1}$ and modifies all kinematical balances \cite{gon97c,gon97d}. Because of the negative value of $\Delta ~E$ \cite{gon97d}, it costs more and more energy, as $E$ increases, to split the incoming longitudinal momentum. The ratio $m^2~c^3~(2~p~\Delta ~E)^{-1}$ between the mass term and the deformation varies like $\sim ~E^{-4}$ , leading to a sharp transition at $E ~\simeq ~ E_{trans}$.

$E_{trans}$ is not the only phenomenological energy scale naturally generated by DRK. In QDRK, the allowed final-state
phase space of two-body collisions is strongly
reduced
\cite{gon97f},
with a sharp fall of partial and total cross-sections,
for cosmic-ray energies above
$E_{lim} ~\approx ~(2~\pi )^{-2/3}~(E_T~a^{-2}~ \alpha ^{-1}~h^2~c^2)^{1/3}$ = $ \alpha ^{-1/3}~(E_T~E_a^2)^{1/3}$,
where $E_T$ is the target energy.
Contrary to $E_{trans}$ that depends on the mass of the particle considered, $E_{lim} $ depends on the target energy $E_T$. Taking for the cosmic background radiation photons an average energy $E_T ~\approx  $ 7 x $10^{-4} ~eV$, we get $E_{lim} ~\approx ~ \alpha ^{-1/3}$ x 5 x $10^{17} ~ eV$ if $E_a$ is the Planck energy. In this case, $\alpha ~\approx  ~10^{-6}$ leads to $E_{lim} ~\approx ~ $ 5 x $10^{19} ~ eV$.

QDRK can generate \cite{gona,gonb} observable phenomena, such as the suppression of the GZK cutoff. Above $E_{trans}$, but also already at $E ~ \simeq ~ E_{trans}$, standard calculations leading of the GZK cutoff do no apply, as the CMB photon energies used can no longer produce the required reactions. For $\alpha ~a^2~>~10^{-72}~cm^2$ ,
and assuming universal values of $\alpha $ and $c$ ,
there is no GZK cutoff for the particles under
consideration \cite{gon97a}. The possible absence of the GZK cutoff would be basically a $E_{lim} $ effect applying to all kinds of particles \cite{gona,gonb}, but it also requires $E ~\gsim ~ E_{trans}$. The existence of the GZK cutoff, if definitely confirmed, will exclude values of $\alpha $ leading to $E_{lim}~ \lsim ~ 10^{20} ~ eV$ and $E_{trans}~ \lsim ~ 10^{20} ~ eV$ in the simple pattern given by (1) $-$ (3). As shown in our 1997 papers and subsequently stressed, \cite{gona,gonb}, the observation of the GZK bound excludes $\alpha ~\gsim ~ 10^{-6} $ for all particles directly concerned by the observation. 

\subsection{Other phenomena}

However, as $E_{trans} ~ \simeq ~$ 2 x $10^{18} ~ eV $ for electrons and positrons with $\alpha ~\simeq ~ 10^{-6} $ and $E_a$ = Planck energy, the flux of cosmic photons in the $\approx ~ 10^{19} ~ eV$ range or above this energy cannot be used for reliable tests of QDRK with the present theoretical uncertainties. For instance, a moderate difference in the value of $\alpha $ between electrons and phonons may lead $\gamma ~ \rightarrow ~ e^+~ e^-$ decays \cite{gon97d}. A detailed study of the fluxes of nucleons and nuclei at ultra-high energy is therefore needed. 

UHECR cross-sections can be altered by LSV through other mechanisms than the change of kinematical balances above $E_{trans}$. One of them would possibly be the effective size of the UHECR in the case of protons or nuclei. Following a simple model developed in \cite{gon97e}, if the relevant inverse Lorentz factor $\gamma ^{-1}$ for a UHE particle of rest size $D$ in the VRF
is corrected by a power series
$\gamma ^{-2}~=~\gamma _R^{-2}~+~\gamma '~\xi ~+~...$ where 
$\gamma _R$ is the usual relativistic Lorentz factor, $\gamma '$ a constant of order 1 and $\xi ~=~\alpha ~(a~\gamma _R)^2~D ^{-2}$, we expect the departure
from standard relativity to play a leading role at 
energies close to
$\approx
~m~c^2~\alpha ~^{-1/4}~(a~D ^{-1})^{-1/2}$ or larger. For a proton, this scale corresponds to $E ~\approx ~
3$ x $10^{19}~eV$ for $a~\approx ~10^{-33}~cm$~
and $\alpha _{proton}~ \approx ~ 10^{-2}$. 
With the same values of the parameters, the maximum velocity 
for a proton occurs at a similar energy. Both energies become $\approx ~
10^{20}~eV$ for $\alpha _{proton}~ \approx ~ 10^{-4}$.

\section{Bounds for QDRK}

For a proton, taking $E_a$ to be the Planck energy and $E_{trans}~ \approx ~ 10^{20} ~ eV$ leads to the value of $\alpha $ : $\alpha _{proton}~ \approx ~ 10^{-6}$, assuming that the GZK cutoff is confirmed and that LSV does not substantially alter the relevant UHECR cross-sections through mechanisms other than kinematics. $\alpha _{proton}~ \approx ~ 10^{-6}$ would then be the upper bound from AUGER-HiRes data if the UHECR were protons. For nuclei made of $N$ nucleons, the upper bound on $\alpha _{Nucleus(N)}$ from the same data becomes $\approx ~ 10^{-6}~ N^2$ using the same equations as for the proton and replacing the proton mass by that of the nucleus. For a carbon nucleus, with the same hypothesis, the upper bound would be $\alpha _{carbon}~ \approx ~ 10^{-4}$. But $\alpha _{carbon}~ \approx ~ 10^{-4}$ is actually the equivalent of $\alpha _{proton}~ \approx ~ 10^{-2}$ \cite{Gonzalez-Mestres2008}. 

More precisely, as discussed in \cite{gon97e}, a high-energy nucleus must be regarded to a first approximation as a set of N nucleons where energy is additive. A simple calculation, assuming $\alpha$ to have the same value for the neutron and for the proton, leads then to $\alpha _{Nucleus(N)} ~ \approx ~ N^{-2} ~ \alpha _{proton}$ in QDRK ($\alpha _{Nucleus(N)} ~ \approx ~ N^{-1} ~ \alpha _{proton}$ in LDRK). Combining this result with that obtained previously, the actual upper bound on $\alpha _{proton}$ would be $\approx ~ 10^{-6}~ N^4$ if the UHECR are nuclei with N nucleons. The lightest component of the UHCR spectrum, if properly identified and analyzed with enough statistics, can possibly provide the most stringent bound. However, it should still be verified that such a component does not result from decays or interactions of heavier nuclei. 

\subsection{Hadrons, quarks and gluons}

The composition of the UHECR spectrum is thus an essential information, as the implications for $\alpha _{proton} $ can vary over at least four orders of magnitude. But a more fundamental issue is that of the proton structure, especially as it is now confirmed by lattice calculations \cite{Durr} that the confined energy provides about 95$\%$ of the proton mass. The question of whether $\alpha _{proton} $ is the real "basic" value of the $\alpha $ parameter, and how close it is to the values of $\alpha $ for the photon and for leptons, must therefore be addressed. Then, if confinement still occurs at ultra-high energies and taking a conservative point of view, the value of $\alpha $ for quarks and gluons, $\alpha _{QG}$ (assuming quarks and gluons have similar values of $\alpha $), is expected to be one or two orders of magnitude larger than $\alpha _{proton}$ \cite{Gonzalez-Mestres2008}. Actually, because of the non-perturbative origin of confinement, the ratio $\alpha _{proton} ~ \alpha_{QG}^{-1}$ can be even smaller. Thus, present data on UHECR are clearly not {\it a priori} incompatible with $\alpha _{QG} ~ \approx ~ 1$ if $a$ is taken to be the Planck length. A better knowledge of the detailed proton structure would help to clarify the situation. Furthermore, even admitting the validity of the concept of unification of all forces at the Planck scale, no basic Physics principle compels the physical fundamental scale to coincide with the unification scale. $a$ can therefore be smaller than the Planck length. Exploring possible values of $E_a$ higher than the Planck energy $E_{Planck}$ is therefore not in real contradiction with the basic ideas of standard particle theory.

\subsection{Other possible effects}

With the range of LSV parameters considered, above some
energy $E_{lim} $ between $\sim ~ 10^{22}$ and $\sim ~ 10^{24}$ $eV$, a cosmic
ray will not deposit most of its energy in the atmosphere
and can possibly fake an exotic event with much less energy \cite{gon97f}.

The suppression of the GZK effect can also be delayed if LSV occurs but is weak at the Planck scale. If the fundamental energy is set to be $\sim ~ 10^3$ times the Planck energy, and using the same values of $\alpha $ as in the previous discussion, one gets $E_{trans}~ \gsim ~$ 3 x $10^{21} ~ eV$. As $\Delta E$ grows like $p^3$, kinematical balances can change between the region 5 x $10^{19} ~ eV$ - $10^{20} ~ eV$ and energies an order of magnitude higher. There are also possible thresholds \cite{gonb} able to generate a delayed suppression of the GZK cutoff manifesting itself above the $ 10^{20} ~ eV$ region. Satellite experiments seem necessary to explore the lowest possible fluxes. 

\section{QDRK and spontaneous decays}

As emphasized in previous papers (e.g. \cite{gon97d}), all stable elementary particles must in principle have the some value of $\alpha $ up to small corrections. Otherwise, those with lower $\alpha $ would undergo spontaneous decays or spontaneously radiate in vacuum. This requirement becomes less stringent if significant bounds can be set on the LSV parameters. If the proton has a smaller value of $\alpha $, it can for instance decay by emitting photons.

Assume $\alpha _{proton} ~ = ~ \Lambda ~ \alpha _{\gamma } $ , with $0 ~ < ~ \Lambda ~ < ~1 $, where $\alpha _{\gamma } $ is the value of $\alpha $ for the photon. A proton with momentum $p$ can spontaneously radiate a photon with momentum $p_1$ if :
\begin{equation} \begin{split}
\alpha _{\gamma } ~ p_1^3 ~ (c ~ E_a^{-1})^2 ~> ~ (m ~ c)^2 ~ [(p ~ - ~ p_1)^{-1} \\ - ~ p^{-1}] ~ + ~ \Lambda  \alpha _{\gamma } ~ [p^3 ~ - ~ (p ~ - ~ p_1)^3] ~ (c ~E_a^{-1})^2
\end{split} \end{equation}
$m$ being the proton mass. Then, a necessary condition for spontaneous decay to be kinematically allowed at high enough energy is \cite{Gonzalez-Mestres2008} :
\begin{equation}
G ~(\zeta) ~ = ~ \zeta ^3 ~ - ~ \Lambda ~ [1 ~ - ~ (1 ~ - ~ \zeta ) ^3] ~> ~ 0
\end{equation}
where $\zeta  ~ = ~ p_1 ~ p^{-1}$, and $\zeta $ must be in the range $0 ~ < ~ \zeta ~ < ~ 1$ . $G ~(\zeta)$ becomes positive at $\zeta ~> ~\zeta _0 ~ = ~ [(12 ~ \Lambda ~ - 3 ~ \Lambda ^2)^{1/2} ~ - ~ 3 ~ \Lambda ] ~ [2 ~ (1 ~ - ~ \Lambda )]^{-1}$. 
The spontaneous decay is kinematically allowed for $\zeta ~ > ~ \zeta _0 $ and :
\begin{equation}
\begin{split}
\alpha _{\gamma }  ~ p{^4} ~ (m ~E_a)^{-2} ~> ~ \zeta ~ [(1 ~ - \zeta ) ~ G ~(\zeta )] ^{-1} ~ \\ = ~ [(1 ~ - \zeta ) ~ (\zeta ^2 ~ - \zeta ^2 ~ \Lambda ~ + 3 ~ \zeta ~ \Lambda ~ - ~ 3 ~ \Lambda )]^{-1}
\end{split}
\end{equation}

Numerical values of $\zeta _0 $ are : $\simeq ~ 0.44 $ for $\Lambda ~ \sim ~ 0.1$ ; $\simeq ~ 0.16 $ for $\Lambda ~ \sim ~ 0.01$ ; $\simeq ~ 0.055 $ for $\Lambda ~ \sim ~ 10^{-3}$ ... For smaller values of $\Lambda $, one has $\zeta _0 ~ \simeq ~ (3 ~ \Lambda )^{1/2}$. The process has no equivalent in special relativity and one may expect unusually small matrix elements for this new kind of decays. A similar mechanism can exist for nuclei, and may be accompanied by spontaneous decay into nucleons or lighter nuclei.
If $\alpha_{\gamma } $ lies in the range $0.1 - 1$, spontaneous photon emission induced by LSV would be allowed for protons or nuclei above some scale between $E ~ \sim ~ 10^{19} ~ eV$ and the GZK energy. But, because of matrix elements and phase space, the phenomenon may manifest itself only at higher energies and/or at very large distance scales. It is therefore not impossible that spontaneous decays of UHECR (protons and nuclei) due to LSV fake the GZK cutoff. 

This mechanism is unrelated to the inhibition of synchrotron radiation in LSV patterns predicted in our 1997-2000 papers and more precisely studied in \cite{gon001}, leading to tests of DRK like that presented in \cite{Jacobson} for a version of LDRK. The UHE photons considered here have much higher energies than those of synchrotron radiation. 

\section{Superbradyons}

Superbradyons \cite{gonSL} would have positive mass and energy, and a critical speed in vacuum much larger than the speed of light. A simple assumption is that they obey their own internal symmetries and a new Lorentz invariance with critical speed $c_s ~\gg ~ c$. They can be the ultimate constituents of matter and of the physical vacuum in a deeper way than standard preon models and feeling different properties of space and time, just as a photons or free nucleons as compared to the phonons in a solid. In a universe made of superbradyons, our standard particles would look to some extent like phonons, solitons or quasiparticles in a body made of conventional matter. They would be excitations of the physical vacuum. 

Superbradyons can directly produce several kinds of observable effects, if they exist inside our standard Universe as individual constituents of the vacuum or as free particles. They would then always have values of $E~p^{-1}$ much larger than $c$ and be in principle able to emit "Cherenkov" radiation in the form of very high-energy standard particles. However, free superbradyons present in our Universe may have lost all their energy available for "Cherenkov" radiation and form a dark matter sea. A "relativistic" superbradyon (speed $v ~ \sim ~ c_s$) would be kinematically allowed to emit conventional particles in vacuum with a characteristic signature related to the property $E~p^{-1}~\gg ~ c$ (back to back or isotropic events). "Nonrelativistic" superbradyons ($v ~\ll ~ c_s$) can also emit "Cherenkov" radiation in vacuum, but only if their speed is larger than $c$. Writing for a "nonrelativistic" superbradyon 
$p ~ \simeq ~ m ~ v$ and :
\begin{equation}
E ~ \simeq ~ m ~ c_s ^2 ~ ~ + ~ m ~ v^2/2
\end{equation}
the energy loss for longitudinally emitting a massless bradyon with momentum $p'$ would be $< ~ p' ~ c$ for $v ~ < ~ c$. Then, spontaneous decay is no longuer allowed as it cannot preserve energy and momentum conservation. Our Universe may thus contain a sea of superbradyons with speeds $\simeq ~ c$. 

If superbradyons have not yet lost all their available energy for spontaneous radiation, they can possibly be detected by experiments like AUGER \cite{gonSL}. A clear identification would in principle be difficult for cosmic rays from superbradyonic decays, but a superbradyonic interaction in the atmosphere would have a clear signature because of the atypical event profile. Satellite experiments should in principle be more sensitive to possible (rare) events from superbradyons. But it may happen that, for basic physical reasons, superbradyons are present near standard cosmic accelerators and radiate from the same places where conventional acceleration occurs.

Superbradyons open the way to a new cosmology, with a new approach to inflation and to the cosmological constant problem as well as new forms of dark matter and dark energy \cite{Gonzalez-Mestres2008,gona,gonb}. While tachyons preserve standard Lorentz invariance, superbradyons explicitly violate it and have a different critical speed in vacuum. Superbradyon cosmology will therefore not involve the same basic mechanisms as the models using tachyons proposed since 2001 \cite{TachyonsCosmo}. Superbradyon distribution and collective interaction properties will also be major issues, as it is already the case in standard physics where the possible role of curvature and matter inhomogeneities in our Universe is not yet fully understood \cite{Buchert} . 

\section{Conclusion and comments}
\vskip 3mm 
\noindent
Commenting on the relativity principle he had formulated in 1895  \cite{Poincarea}, Henri Poincar\'e wrote in 1901 \cite{Poincareb} : 
{\it "This principle will be confirmed with increasing precision, as measurements become more and more accurate"}.
Twenty years later, Albert Einstein wrote in "Geometry and Experiment" \cite{Einstein}:
{\it "The interpretation of geometry advocated here cannot be directly applied to submolecular spaces... it might turn out that such an extrapolation is just as incorrect as an extension of the concept of temperature to particles of a solid of molecular dimensions"}. 
Poincar\'e's prediction remains experimentally valid, but the word of caution by Einstein has not yet been confirmed by data at the highest available energies. This is, today, the most fundamental problem of Particle Physics and Astrophysics. 

The essential issue of the composition of the UHECR cosmic-ray spectrum needs further clarification with better statistics in current experiments. Whatever the result, AUGER and Hires data are yielding bounds and constraints of LSV models and basic parameters. But another important range will remain to be tested experimentally. QDRK provides an example of models that AUGER and Hires results will not allow to exclude, even if a range of parameters will be ruled out by the data. Taking quarks and gluons as the relevant elementary particles to build the QDRK pattern for hadrons, and assuming the UHECR are intermediate mass nuclei, the actual strength of LSV can be close to 1 at the Planck scale without conflicting with AUGER and HiRes results. If further data and analysis allow to clearly identify a UHECR proton component, the origin of such protons will still have to be elucidated before drawing any theoretical conclusion. 

LSV can also generate effects similar to the GZK cutoff but with a completely different dynamical origin, through mechanisms that appear difficult to rule out with present experiments. 

Obviously, the experimental study of possible Lorentz symmetry violations at ultra-high energy must be pursued beyond HiRes and AUGER, including satellite experiments, in order to further explore Physics as close as possible to the Planck scale. Rather than closing a debate on the GZK cutoff, AUGER results are likely to open new ways for research on UHECR and on the basic Physics they can unravel. Two main classes of LSV scenarios must be considered : 

- {\it Lorentz symmetry violation within present standard particle physics}. LSV modifies present particle physics theory, without fundamentally changing it. It may originate from quantum gravity or from another dynamical phenomenon, at the Planck scale or at some fundamental scale not far from the Planck scale or beyond it. HiRes and AUGER data will play an important role in constraining models and parameters, but many possibilities will still be left open.

- {\it Fundamentally new physics}. Standard particle physics and cosmology is not the ultimate theory. The particles of the present standard theory are not necessarily the elementary constituents of matter. $c$ may not be the ultimate critical speed in vacuum. Our Universe may then be a bubble in superbradyonic matter \cite{gonSL}, with our vacuum as the ground state, $a$ as a correlation length and the standard particles as excitations \cite{gon97a}. Worries about the cosmological constant and dark energy would hopefully be solved in this way.

Globally, many fundamental pieces of information are still missing, and require a much larger amount of data at all explorable energies above the $\sim ~ 10^{19} eV$ region. Algorithms allowing to precisely estimate the energy of each event from the available data are another important issue at such extreme energies. A better understanding of the theoretical and phenomenological implications of LSV is also necessary, including its effects on UHECR interactions with the atmosphere or with the CMB radiation photons. 

Further experiments and theoretical developments are therefore required in order to complete the work of the HiRes and AUGER collaborations as well as existing theoretical models, and eventually open the way to a full understanding of physics at the Planck scale or beyond it.

\end{document}